*2024-2025 CRA Quadrennial Paper*

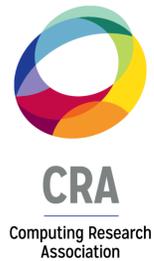

# Reclaiming the Future: American Information Technology Leadership in an Era of Global Competition

Alex Aiken (Stanford University), David Jensen (University of Massachusetts Amherst), Catherine Gill (CRA), William Gropp (University of Illinois Urbana-Champaign), Peter Harsha (CRA), Brian Mosley (CRA), Daniel Reed (University of Utah), William Regli (University of Maryland, College Park)

**The United States risks losing its global leadership in information technology research due to declining basic research funding, challenges in attracting talent, and tensions between research security and openness.**

For decades, the United States has benefited from federal government investments in fundamental research across the sciences and engineering, particularly in computing and other forms of information technology (IT). The nation's research ecosystem — rooted in a unique partnership among the federal government, academic institutions and their researchers, and the private sector — has supported the generation of new knowledge and the creation of new industries. It is well documented that many technological breakthroughs that have become mainstays of the United States' national and economic security — including microchips, the Internet, and most recently, artificial intelligence — were initially funded by the federal government. While the U.S. research ecosystem is envied worldwide, particularly in computing and IT research, its continued success is not guaranteed.

In this paper, we outline multiple areas where the country's research efforts are faltering and U.S. leadership is at risk — with profound implications for our global competitiveness and national security — and suggest key actions that Congress and the Executive Branch should take to arrest and reverse this decline. Most, if not all, of our observations are not new. The current situation is the result of long-term trends that have been well-documented for years and even decades.

What gives this matter new urgency is that the nature of our global competition has changed. Recognizing the importance of technology to its ambitions, China has openly stated its goal to become the world's preeminent scientific and technological power, targeting critical technologies such as artificial intelligence and quantum computing. More tellingly, China is making the investments needed to reach this goal and is the first rival in nearly a century to present a serious challenge to the scientific and technological leadership of the United States.

The key research policies that the U.S. has followed since the end of WWII have been copied by our competitors because of the outstanding track record of these policies in delivering national prosperity. Meanwhile, we are quietly allowing these practices to atrophy.

We point to three core areas of concern:

## 1. Basic Research Funding

Funding for basic research in the United States, as a fraction of national wealth, has been in decline for decades. In 2019, the most recent year with data [compiled by AAAS](), federally supported basic R&D was 0.20 percent of GDP, the lowest level since 1980. The budgetary situation since 2019 has also become unstable. As an example, the budget for the U.S. National Science Foundation (NSF) saw [a historic budget increase]() in Fiscal Year 2023 and then a cut in Fiscal Year 2024, reversing most of those gains in a single year. As another example from further back, one analysis concluded that the [Budget Control Act era]() (2011-2021) reduced overall growth in investment in R&D by $200 billion during that time period.

These budget outcomes have occurred despite overwhelming evidence of the return on investment for fundamental research. [By one measure](), the return on investment of non-defense government R&D is between 150 and 300 percent. As with all basic research in the sciences, investments in research in computing and IT have lead times that span years or decades, but when they pay off, they often do so spectacularly and in unexpected ways.

Industry and academic research play complementary roles: New ideas are often first articulated and nurtured in academia and are picked up in industry only when an eventual path to profitable products becomes reasonably clear. Thus, the basic research investments made by the federal government provide the raw material of new ideas, techniques, and trained people that industry eventually turns into new products and occasionally entirely new industries. This interaction between federal funding, academia, and industry, and between fundamental research and eventual products, has been documented repeatedly by the National Academies' Computer Science and Telecommunications Board [Tire Tracks chart]().

Internationally, the United States has seen its advantages in research erode, and we are in danger of falling behind our competitors. China surpassed the United States in the number of [highly cited computer science publications]() a decade ago and now files twice as many patent applications across all fields as the U.S. every year. When Congress approved the cuts to research agencies in the Fiscal Year 2024 budget, the People's Republic of China [announced guaranteed increases]() to its fundamental research programs.

Why, then, is the sky not falling? The United States continues to dominate most aspects of information technology in the marketplace, and recent breakthrough innovations, such as Large

Language Models, have happened in the West, not in China. What policy makers need to understand is that metrics such as national research budgets, papers written, and patents filed are leading indicators of research prowess, whereas marketplace success and economic growth are lagging indicators of past research investments. The United States' current position in AI is the result of research investments made in the 1980's and 1990's that laid the foundation of key ideas and built a community of researchers seeking to realize those ideas in practice. Our greatest danger is failing to recognize that the research investments we make today will have their greatest impact 20 to 30 years from now — and that if we do not make those investments while our competitors do, we will be ceding the future to them.

In this light, the erratic computing research budgetary and policy direction should be corrected. Consistent, sustainable, and growing budgets must be pursued for the key federal research agencies: NSF, the Department of Energy (DOE) Office of Science, the National Institute of Standards and Technology (NIST), NASA, the National Institutes of Health (NIH), and the Defense research accounts. Coupled with more funding for research, key pieces of authorization legislation, such as the CREATE AI Act and the NSF AI Education Act of 2024, must be passed into law. Funding and legislative backing are key features of maintaining a successful national research strategy.

## 2. Attracting Talent

U.S. research institutions, and particularly U.S. universities, have been a magnet for the best and brightest around the world for generations. International students come to the U.S., are educated at U.S. universities, and then often stay and contribute in myriad ways to U.S. academia, industry, and government. The openness and attractiveness of the U.S. system of research education to international students has been a key factor in maintaining the pre-eminence of U.S. research. While hard data is difficult to obtain, anecdotal evidence suggests that some international students are now subject to much more scrutiny, and it is not uncommon to hear of cases of students who are unable to obtain a student visa after being admitted to a U.S. university. While concerns about research security (discussed further below) are legitimate, the federal government and academia need to work together on policies that address the research security issues while continuing to attract outstanding talent.

Conversely, the number of domestic students remains problematic. Relative to the U.S. population, fewer domestic students pursue research careers than in some of our competitor countries. Historically, as countries develop economically and establish their own science and technology sectors, fewer of their students come to the U.S. to study. To sustain a thriving research enterprise in the long term, the U.S. must continue attracting the very best international students *and* increase the number of qualified domestic students pursuing advanced degrees. Encouraging more undergraduates, especially in computing, to continue on to graduate school is one of the most effective ways to boost domestic participation in research careers. The United States should aspire to be both the destination for the best and brightest minds globally and a deep source of domestically cultivated talent.

## 3. Openness in Research

It is well-known that certain countries and foreign organizations try to exploit the openness of the U.S. research ecosystem to gain an unfair competitive advantage. The academic community takes the matter seriously, and American universities have been [taking actions for years](#) to protect the outcomes of the nation's research efforts.

At the same time, it is important to understand that adding restrictions to the openness of the research enterprise risks damaging the very system that we want to protect. Science advances most quickly in open environments where researchers can easily collaborate; conversely, when restrictions are placed on how scientists can interact, it inevitably slows the rate of progress.

Thus, the country must strive to maintain a balance between protecting its research investments from foreign exploitation, and maintaining and encouraging the collaboration and open exchange of ideas that makes the U.S. research ecosystem so uniquely productive. In particular, the rules around research security should be transparent and uniformly applied. Risk assessments should be limited to objective criteria and exclude retrospective judgments of practices that were legal and even encouraged just a few years ago. Additionally, special care must be applied to avoid any targeting based on racial and ethnic backgrounds. New regulations should account for compliance costs and whether research institutions might forgo working in certain areas or on certain types of projects due to compliance burdens. Finally, federal research agencies should strive to reach a stable point as quickly as practicable where research security regulations are no longer in flux; this is the only point at which the U.S. research enterprise will be able to fully adapt to these requirements.

Congress and the next Administration need to reaffirm the policy articulated in [NSDD-189](#): namely that, "the products of fundamental research remain unrestricted…to the maximum extent possible," and, "the mechanism for control of information generated during federally-funded fundamental research in science, technology and engineering at colleges, universities and laboratories is classification." A new "sensitive but not classified" or equivalent designation would only hinder the flow of fundamental research findings and limit the potential impact of the nation's research investments.

As concluded in the 2019 report ["Fundamental Research Security,"](#) authored by JASON, "many of the problems of foreign influence that have been identified are ones that can be addressed within the framework of research integrity, and that the benefits of openness in research and of the inclusion of talented foreign researchers dictate against measures that would wall off particular areas of fundamental research."

## Recommendations

We recommend the following actions:

1. **Fully fund the research agencies' authorizations contained in the CHIPS & Science Act of 2022.** At a minimum, the research agencies listed in the CHIPS Act should have their funding restored to FY23 levels and be exempt from any automatic or targeted funding cuts enacted in the future.

2. **Address the need for more domestic students who pursue advanced degrees in computer science and related fields.** Each federal research agency should be tasked with exploring the needs and costs of establishing an undergraduate-specific cooperative research program aligned with their mission. Additional and dedicated funding for this type of workforce development program would be essential, as agencies currently lack the budgets to fully integrate undergraduates into the research enterprise.

3. **Redesign visa policies for foreign students with the explicit goal of attracting the best and brightest while ensuring research security.** A strong example of such a proposal is to fast-track green cards for individuals who earn a STEM PhD at a U.S. institution. Additionally, to make the country as welcoming as possible for international talent, the immigration process for those seeking advanced degrees in STEM fields should be streamlined and simplified.

4. **Reaffirm the principles of NSDD-189 to ensure fundamental research remains unrestricted to the fullest extent possible.** Congress should pass, and the next President should sign into law, a sense of Congress resolution reaffirming the policies outlined in NSDD-189, emphasizing that fundamental research should remain unrestricted to the maximum extent possible.

---

*This quadrennial paper is part of a series compiled every four years by the **Computing Research Association (CRA)** and members of the computing research community to inform policymakers, community members, and the public about key research opportunities in areas of national priority. The selected topics reflect mutual interests across various subdisciplines within the computing research field. These papers explore potential research directions, challenges, and recommendations. The opinions expressed are those of the authors and CRA and do not represent the views of the organizations with which they are affiliated.*2024-2025 CRA Quadrennial Paper    5